\documentclass[aps,twocolumn,nofootinbib,showpacs, prd]{revtex4-2}

\usepackage{hyperref}
\usepackage{amsmath,amsfonts,amssymb}
\hypersetup{dvips,dvipdfm,colorlinks=true,urlcolor=magenta,filecolor=magenta,linktoc=page,citecolor=red,linkcolor=blue,bookmarks=true}
\usepackage{graphicx,epstopdf}
\usepackage{multirow}
\usepackage{natbib}
\begin{document}
	\title{Raychaudhuri Equation and Weyl-Driven Shear: A Weak-Field Approach to Lensing and Gravitational Waves}
	\author{Madhukrishna Chakraborty} \email{chakmadhu1997@gmail.com/\\madhukrishna.c@technoindiaeducation.com}
	\author{Subenoy Chakraborty}\email{schakraborty.math@gmail.com (corresponding author)}
	\affiliation{*Department of Mathematics, Techno India University, Kolkata-700091, West Bengal, India\\
		$^\dag$Department of Mathematics, Brainware University, Kolkata-700125, West Bengal, India\\
		$^\dag$Shinawatra University, 99 Moo 10, Bangtoey, Samkhok, Pathum Thani 12160, Thailand\\
		$^\dag$INTI International University, Persiaran Perdana BBN, Putra Nilai, 71800 Nilai, Malaysia}
	\begin{abstract}
		The letter studies phenomena like gravitational wave propagation and gravitational lensing using the celebrated Raychaudhuri equation (RE) in the weak field limit. Newtonian analogue of Relativistic RE has been explored. In doing so, role of shear has been found to be extremely important in explaining these phenomena. Consequently, the RE for shear has been used in course of the study and importance of Weyl curvature tensor in lensing and gravity wave propagation has been explicitly shown using a damped harmonic oscillator approach.
	\end{abstract}
	\maketitle
	\section{Introduction}
	In the framework of general relativity \cite{Wald:1984rg},  the Raychaudhuri Equation (RE) plays a pivotal role in characterizing the evolution of a congruence of geodesics which are essentially bundles of nearby worldlines  within a curved spacetime \cite{Chakraborty:2023ork},\cite{Kar:2006ms},\cite{Chakraborty:2024obs}. First derived by Amal Kumar Raychaudhuri in the 1950s, this geometric identity underpins the celebrated singularity theorems of  Penrose and Hawking \cite{Penrose:1964wq}, \cite{Hawking:1970zqf}. The RE governs the rate of change of the expansion scalar ($\theta$), which quantifies whether geodesics are converging or diverging, and encapsulates how shear ($\sigma$), vorticity ($\omega$), and spacetime curvature interact to influence the evolution of these geodesic congruences and the equation has an influence in cosmology \cite{Chakraborty:2024wty}. For both timelike and null congruences, the REs reveal the dynamics of gravitational focusing. They demonstrate how geodesics are distorted by shear (anisotropic deformations), twisted by vorticity (rotation), and focused by the Ricci curvature linking geometry with the energy-momentum content via Einstein's equations. This equation is not merely of theoretical interest but find application in a variety of physical scenarios, from gravitational collapse and black hole formation to cosmological dynamics \cite{Chakraborty:2024wty}.
	
	Two such physical phenomena where the RE offers profound insight are gravitational lensing \cite{Bartelmann:2010fz} and gravitational waves \cite{Barausse:2023oov}. Gravitational lensing \cite{AbhishekChowdhuri:2023ekr} refers to the bending and distortion of light as it propagates through curved spacetime, typically influenced by the presence of mass. It is fundamentally a consequence of the focusing of null geodesics, precisely what the null RE describes. On the other hand, gravitational waves are ripples in spacetime caused by time-varying mass quadrupole moments \cite{LIGOScientific:2017vwq}. These influence the propagation of geodesics by inducing oscillatory tidal deformations, those are inherently tied to the Weyl curvature tensor and captured by the evolution of the shear.
	While gravitational lensing is usually attributed to matter-induced curvature (via the Ricci tensor), and GWs arise from vacuum spacetime distortions (encoded in the Weyl tensor), both phenomena can be addressed through the lens of shear evolution within the RE framework. This convergence offers an intriguing pathway to unify these seemingly disparate effects under a single geometric identity.
	The motivation for connecting the RE with gravitational lensing and gravitational wave dynamics arises from their common dependence on the geodesic deviation of null congruences and the shear tensor that governs their distortion. In this letter we  explore how, in the weak-field limit, both GWs and weak lensing can be effectively described by the null RE and particularly by the evolution of shear in a twist-free (vorticity-free) congruence. In this approach, gravitational waves, which in vacuum do not contribute to the Ricci curvature, still influence the distortion of geodesics via the Weyl tensor and this is manifested as shear in the congruence. Similarly, weak gravitational lensing is observed when light propagates through nearly empty or mildly curved regions of space. Interestingly, this is also governed by changes in shear rather than expansion. The letter introduces a damped harmonic oscillator analogy to describe how the shear evolves under the influence of the Weyl curvature, thereby capturing both GW propagation and lensing distortion in a unified mathematical form.
	
	This perspective not only provides a covariant, gauge-invariant characterization of these phenomena but also reveals the same geometric mechanism, namely the evolution of shear encoded in the RE. The motivation, therefore is to uncover and emphasize this deep geometric interplay thereby offering a richer understanding of GW amplitudes and lensing distortions through a common theoretical lens.

		Before proceeding, it is useful to clarify the scope and approximations adopted in this work. The analysis is performed within the framework of standard General Relativity using the Raychaudhuri equation as the central geometric tool describing the evolution of geodesic congruences. We primarily consider the weak-field regime where the spacetime metric can be written as a small perturbation around the Minkowski background and matter velocities are assumed to be non-relativistic. Under these conditions the Einstein equations reduce to their Newtonian limit, allowing a transparent connection between the relativistic Raychaudhuri formalism and classical gravitational dynamics. When discussing gravitational waves and lensing, we allow for dynamical perturbations of spacetime but still remain within the weak-curvature approximation relevant for astrophysical observations far from strong-field sources. These assumptions enable us to highlight the geometric role of shear in a simplified yet physically meaningful setting.

	The layout of the letter is: Section II gives an overview of RE in general relativity, Section III deals with the Newtonian version of the relativistic RE, Section IV explains phenomena like GW and lensing using null RE via Damped harmonic oscillator approach. The letter ends with concluding remarks in Section V.

	\section{The Raychaudhuri equation in General Relativity}
	The RE characterizes the kinematics of flows in a geometrical space. The equation plays a fundamental role in general relativity by describing how a bundle (congruence) of nearby geodesics evolves as it moves through spacetime. Precisely, the equation dictates how the bundle of geodesics are focused, sheared and / or twisted in space-time as it evolves. This phenomenon can be described using a precise mathematical expression for the rate of change of the expansion scalar (denoted by $\theta$), which measures the divergence or convergence of geodesics in the congruence.

	 We consider a family of fundamental observers moving along a time-like 4-velocity vector field, say $u_{a}$. The symmetric space-like (induced) metric tensor 
		\begin{equation}
			\gamma_{ab}=\eta_{ab}+u_{a}u_{b}
		\end{equation}
		satisfies the orthogonality condition ,$u^{a}\gamma_{ab}=0$. Define,
		\begin{equation}
			D_{a} =\gamma_{a}^{b}{~}\nabla_{b}
		\end{equation} operating in the 3-D hyper-surface. The kinematics of the fundamental observers are expressed by four irreducible variables, found post decomposition of the gradient of $u_{a}$, the velocity vector field as
		\begin{equation}
			B_{ab}=\nabla_{b}u_{a}=\dfrac{1}{3}\theta{~} \gamma_{ab}+\sigma_{ab}+\omega_{ab}-A_{a}u_{b}
		\end{equation}
		where $B_{ab}$ is called the deformation tensor,
		\begin{equation}\label{eq5*}
			\theta= B^{a}_{a}=\nabla_{a}u^{a}
		\end{equation}
		is the volume scalar/ expansion scalar (trace part of $B_{ab}$) .
		\begin{equation}\label{eq6*}
			\sigma_{ab}=\nabla_{(ba)}u-\dfrac{1}{3}\gamma_{ab}\theta
		\end{equation} is the shear tensor (symmetric i.e $\sigma_{ab}=\sigma_{ba}$, traceless i.e $\sigma^{a}_{a}=0$).  Here, $\nabla_{(ba)}u=\nabla_{b}u_a$+$\nabla_{a}u_b$.
		\begin{equation}
			\omega_{ab}=D_{[ab]}u
		\end{equation}
		is the vorticity tensor (anti-symmetric i.e $\omega_{ab}=-\omega_{ba}$).   Here, $D_{[ba]}u=\nabla_{b}u_a$-$\nabla_{a}u_b$.
		\begin{equation}
			A_{a}=u^{b}\nabla_{b}u_{a}
		\end{equation} is the 4-acceleration vector field.

	The REs for a congruence of time-like and null geodesic are given by \cite{Chakraborty:2024tal}
	\begin{eqnarray}
		\dfrac{d\theta}{d\tau}=-\dfrac{\theta^{2}}{3}-2\sigma^{2}+2\omega^{2}-R_{ab}u^{a}u^{b}\label{eq1}\\
		\dfrac{d\theta}{d\lambda}=-\dfrac{\theta^{2}}{2}-2\sigma^{2}+2\omega^{2}-R_{ab}k^{a}k^{b}\label{eq2}
	\end{eqnarray}
	$\theta$ is called the expansion scalar which measures the relative expansion or contraction of the bundle ; $\sigma$ is the shear scalar that measures the kinematic anisotropies of the deformable medium or in other words, it measures how much the bundle is sheared in the course of its evolution ; $\omega$ is called the vorticity scalar that measures the rotation of the bundle i.e, how much the bundle is twisted during evolution ; $R_{ab}$ is the $(0,2)$ Ricci tensor that encodes curvature of the space-time manifold under consideration ; $u^{a}$ is a time-like vector field (applicable for time -like RE given by equation (\ref{eq1})) ; $k^{a}$ is the null vector field (applicable for Null RE given by equation (\ref{eq2})) ; $R_{ab}u^{a}u^{b}/R_{ab}k^{a}k^{b}$ is the Raychaudhuri scalar that plays a very crucial role in geodesic focusing ; $\tau$ is the proper time that becomes the cosmic time in cosmology ; $\lambda$ is the affine parameter. The evolution equations (along the flow) of the quantitites that characterise the flow in a given background spacetime, are essentially the Raychaudhuri equations. Historically speaking, it is the equation for one of the quantitites (the expansion), which is termed as the RE. However, there are similar evolution equations for shear and vorticity in case of time-like (given by equations (\ref{eq3}), (\ref{eq4})) and null geodesics (given by equations (\ref{eq5}), (\ref{eq6})) respectively given by \cite{Kar:2006ms}
	\begin{eqnarray}
		\dfrac{d\sigma_{ab}}{d\tau}=-\dfrac{2}{3}\theta\sigma_{ab}-\sigma_{ac}\sigma^{c}_{b}+\dfrac{1}{3}h_{ab}\left(\sigma^{2}-\omega^{2}\right)+C_{cbad}u^{c}u^{d}\nonumber\\
		+\dfrac{1}{2}\left(h_{ac}h_{bd}R^{cd}-\dfrac{1}{3}h_{ab}h_{cd}R^{cd}\right)\label{eq3}
	\end{eqnarray}
	\begin{eqnarray}
		\dfrac{d\omega_{ab}}{d\tau}=-\dfrac{2}{3}\theta\omega_{ab}-2\sigma^{c}_{[b^{\omega}a]c}\label{eq4}
	\end{eqnarray}
	\begin{eqnarray}
		\dfrac{d\sigma_{ab}}{d\lambda}=-\theta\sigma_{ab}-\sigma_{ac}\sigma^{c}_{b}+\dfrac{1}{2}h_{ab}\left(\sigma^{2}-\omega^{2}\right)+C_{cbad}k^{c}k^{d}\nonumber\\
		+\dfrac{1}{2}\left(h_{ac}h_{bd}R^{cd}-\dfrac{1}{2}h_{ab}h_{cd}R^{cd}\right)\label{eq5}
	\end{eqnarray}
	\begin{eqnarray}
		\dfrac{d\omega_{ab}}{d\lambda}=-\theta\omega_{ab}-2\sigma^{c}_{[b^{\omega}a]c}\label{eq6}
	\end{eqnarray}
	Here, $\sigma^{2}=\sigma_{ab}\sigma^{ab},~\omega^{2}=\omega_{ab}\omega^{ab}$ and $C_{cbad}$ is the Weyl tensor. One must realise that these are not equations but identities often named as Raychaudhuri identities or Codazzi-Raychaudhuri identities. However, the identities  become equations once the Einstein field equations or any other geometric property (e.g. vacuum or Einstein space, etc.)  are used as an extra input. However, we shall continue to use the term equations in this letter.

	Although, RE primarily finds an immediate application in studying Singularity theorems by Penrose and Hawking but its wider applicability in geometry, mechanics and other allied fields is indisputable. In General Relativity, RE leads to the Focusing Theorem (FT) that paved the way for the seminal singularity theorems by Hawking and Penrose. In Einstein gravity, if we consider usual matter (a matter satisfying the Strong Energy Condition) then 
	\begin{equation}
		R_{ab}-\dfrac{1}{2}Rg_{ab}=T_{ab}
	\end{equation} 
	gives 
	\begin{equation}
		R_{ab}u^{a}u^{b}\geq 0 / ~	R_{ab}k^{a}k^{b}\geq 0
	\end{equation}
	Therefore, RE for time-like and null geodesic congruence can be written as 
	\begin{eqnarray}
		\dfrac{d\theta}{d\tau}+\dfrac{\theta^{2}}{3}\leq 0\\
		\dfrac{d\theta}{d\lambda}+\dfrac{\theta^{2}}{2}\leq 0
	\end{eqnarray}
	Integrating the above inequalities we get,
	\begin{eqnarray}
		\dfrac{1}{\theta(\tau)}\geq \dfrac{1}{\theta_{0}}+\dfrac{\tau}{3}\\
		\dfrac{1}{\theta(\lambda)}\geq \dfrac{1}{\theta_{0}}+\dfrac{\lambda}{2}
	\end{eqnarray}
	This is the mathematical version of the FT. The theorem states that an initially converging ($\theta_{0}<0$) congruence of time-like / null geodesics develop a caustic within some finite value of the proper time $\tau$ / affine parameter ($\lambda$), leading to congruence singularity\cite{Chakraborty:2023yyz}. It is to be carefully noted that, this caustic may or may not be a curvature singularity. However, focusing along with some global assumptions on matter and space-time geometry might sometimes lead to cosmological and black-hole singularity namely the Singularity Theorems. Later, it was pointed out by Landau \cite{Landau:1975pou} that focusing alone can not imply singularity but if there is singularity then focusing is inevitable. Therefore RE tells us that gravity tends to focus geodesic bundles, potentially leading to caustics or singularities as the case may be. It describes the competition between expansion, shear, rotation, and spacetime curvature, making it central to understanding gravitational collapse, singularities, lensing, and wave propagation in curved spacetimes (gravitational waves). There is enough literature on the implications of RE in identifying or mitigating both black-hole as well as cosmological singularities like \cite{Chakraborty:2024khs}. However, this letter studies phenomena like gravitational lensing and gravity wave propagation via the celebrated RE.
	\section{Newtonian analogue of Raychaudhuri equation: Derivation and Interpretation}
	In this section, we deduce a Newtonian analogue of relativistic RE. The Newtonian limit of the RE arises when considering the behavior of a congruence of worldlines in a weak gravitational field and slow-moving matter, where relativistic effects are negligible. In this limit, the equation reduces to a form that describes the evolution of the volume expansion of a fluid element under classical Newtonian gravity. 
		In the Newtonian limit the Raychaudhuri equation reduces to a relation connecting the divergence of the velocity field with the local matter density and tidal gravitational forces. Using the linearized Einstein equations, the curvature term can be written in terms of the Newtonian gravitational potential $U$, which satisfies the Poisson equation
		\begin{equation}
			\nabla^2 U = 4\pi G\rho .
		\end{equation}
		In Newtonian fluid dynamics the evolution of the velocity field is governed by the Euler equation
		\begin{equation}
			\frac{d\vec{v}}{dt} = - \nabla U ,
		\end{equation}
		while the divergence of the velocity field appears naturally in the continuity equation describing mass conservation. The Newtonian limit of the Raychaudhuri equation therefore provides a geometric interpretation of these classical relations by expressing the divergence of the flow in terms of the gravitational potential sourced by the matter density. For the mathematical reduction, see the Appendix.
	
	We consider the time-like and null RE (\ref{eq1}) and (\ref{eq2}). We further consider Einstein's field equations to write 
	\begin{eqnarray}
		R_{ab}u^{a}u^{b}=T_{ab}u^{a}u^{b}+\frac{1}{2}T\\
		R_{ab}k^{a}k^{b}=T_{ab}k^{a}k^{b}
	\end{eqnarray}
	In linearized gravity,
	\begin{equation}
		g_{ab}=\eta_{ab}+h_{ab}, ~|h_{ab}|<<1\label{eq15}
	\end{equation}
	and $\eta_{ab}$ is the usual Minkowskian metric. In Newtonian limit, the Einstein field equations reduce to Poisson equation for Newtonian gravity i.e, 
	\begin{equation}
		\nabla^{2}h_{00}=-\dfrac{8\Pi G\rho}{c^{2}}, ~h_{00}=-\dfrac{2U}{c^{2}}
	\end{equation}
	$U$ is the Newtonian gravitational potential. Therefore, we have 
	\begin{equation}
		\nabla^{2}U=4\Pi G\rho\label{eq17}
	\end{equation}
	For perfect fluid, 
	\begin{eqnarray}
		R_{ab}u^{a}u^{b}=(\rho+3p)=\rho(1+3w)\\
		R_{ab}k^{a}k^{b}=(\rho+p)=\rho(1+w)
	\end{eqnarray}
	In deriving Newtonian limit from GR we have two basic assumptions: (i) geometry is deviated from Minkowskian by first order given by (\ref{eq15}) and (ii) all quantities do not vary significantly w.r.t time $t$. 
		The approximation that all quantities vary slowly with respect to the coordinate time $t$ corresponds to the quasi-static Newtonian limit of General Relativity. In this regime, the metric perturbations $h_{ab}$ and matter variables evolve on timescales much longer than the characteristic dynamical time of the congruence, allowing time derivatives of these quantities to be neglected to leading order. The parameters $\tau$ and $\lambda$ used in the Raychaudhuri equations denote respectively the proper time along time-like geodesics and the affine parameter along null geodesics. Neglecting explicit time derivatives of the background quantities does not eliminate the dependence of the kinematic quantities $\theta$, $\sigma$ and $\omega$ on $\tau$ or $\lambda$. Rather, it implies that the background spacetime is approximately stationary on the timescale considered. Consequently the focusing theorem remains valid since its derivation rely on the geometric properties of the congruence and the energy conditions, which are unaffected by this quasi-static approximation. Under these assumptions the time-like and null RE becomes
	\begin{eqnarray}
		2\sigma^{2}-2\omega^{2}=-\dfrac{(1+3w)}{4\Pi G}\nabla^{2}U\label{eq20}\\
		2\sigma^{2}-2\omega^{2}=-\dfrac{(1+w)}{4\Pi G}\nabla^{2}U\label{eq21}
	\end{eqnarray}
	This is because, $\frac{d\theta}{d\tau}$ or $\frac{d\theta}{d\lambda}$ vanishes and also $\theta=\frac{V'}{V}$ vanishes where $V$ is the volume and `$'$' represents differentiation w.r.t $\tau/\lambda$ as the case may be. Equations (\ref{eq20}) and (\ref{eq21}) are analogous to relativistic time-like/ null RE in Newtonian limit.
	Therefore, the Newtonian analogue of relativistic RE is a  geometric identity which encodes a geometric quantity $U$, the gravitational potential in terms of the kinematic quantities $\sigma$ and $\omega$. For hyper-surface orthogonal ($\omega=0$) congruence of time-like/ null geodesics only shear determines the gravitational potential.

	It is crucial to note that the formulation adopted here remains entirely consistent with standard General Relativity and reproduces the expected results in the appropriate limits. The Raychaudhuri equation so used throughout this work is a geometric identity in the Lorentzian geometry. Therefore, it holds in any spacetime described by Einstein's theory. The weak-field approximation is implemented through the linearized metric expansion (\ref{eq15}) which corresponds to the conventional perturbative limit of General Relativity. Under this approximation, Einstein's field equations reduce to the Poisson equation (\ref{eq17}) thereby recovering the standard Newtonian gravitational potential. Consequently, the relations obtained in equations (\ref{eq20})  and (\ref{eq21}) represent the weak-field limit of the relativistic Raychaudhuri equation. This demonstrates that the present formulation does not modify General Relativity but rather provides an alternative geometric interpretation of gravitational phenomena such as gravitational lensing and gravitational wave propagation through the evolution of the shear tensor within the Raychaudhuri framework.
	\section{Gravitational Waves and Lensing from a Raychaudhuri Perspective}
	In this section we work within the weak-field approximation of General Relativity and consider null geodesic congruences in nearly vacuum regions of spacetime. Unlike the quasi-static approximation used in the Newtonian discussion, time dependence is now allowed since gravitational waves represent dynamical perturbations of the spacetime curvature. However, we assume that the lensing is weak so that the expansion scalar is small, $\theta \approx 0$, which implies that the term $\theta^2$ appearing in the Raychaudhuri equation is negligible compared with the shear contribution.
	Null RE dictates how null geodesic congruence / light rays are focused, sheared or twisted in space-time \cite{Chakraborty:2024dpx}.  
	
	Although, both lensing and gravity waves generate from different physical causes (matter vs space-time perturbation) but can be studied using the null RE. However, we need to consider twist free congruence ($\omega=0$) to ensure that light rays are emitted from a point source or received by a distant observer. Also, $2\omega^{2}$ when dropped in RE makes it simpler to comprehend.
	
	Gravitational waves (GWs) are transverse traceless distortions in space-time curvature. Their physical effect is to cause tidal deformation. GWs have either zero or very negligible contribution of Raychaudhuri scalar as $R_{ab}\approx 0$. Therefore, from the evolution of twist free congruence it is clear that shear affects the expansion/ contraction of the congruence. In the context of GWs, it is the Weyl tensor and not the Ricci tensor that comes into picture. Weyl tensor explicitly does not appear directly in RE but influences the shear. It is responsible for tidal distortions in vacuum including GWs. GWs can mimic lensing-like distortion even in vacuum via shear. Therefore, RE dictates that only shear determines the evolution of null congruence, GWs and weak-lensing. In other words, when light rays/ null rays propagate in vacuum or nearly vacuum regions both GWs and tidal lensing effects are prominent in understanding geodesic convergence via null RE and the key role is played by shear. This motivates us to invoke the evolution equation for shear (\ref{eq5}) to study the phenomena like lensing and GWs. In the weak field limit ($R^{cd}=0$), equation (\ref{eq5}) for twist-free congruence reduces to
	\begin{equation}
		\dfrac{d\sigma_{ab}}{d\lambda}=-\theta\sigma_{ab}-\sigma_{ac}\sigma^{c}_{b}+\dfrac{1}{2}h_{ab}\sigma^{2}+C_{cbad}k^{c}k^{d}\label{eq22}
	\end{equation}
	Consider a mode/ single component of shear by $\sigma\approx \sigma_{ab}e^{a}e^{b}$ so that (\ref{eq22}) becomes 
	\begin{equation}
		\frac{d\sigma}{d\lambda}=-\theta\sigma-2\sigma^{2}+C_{acbd}k^{c}k^{d}\label{eq23}
	\end{equation}
	Consider the damping transformation, 
	\begin{equation}
		\sigma(\lambda)=\Sigma(\lambda) e^{-\int \frac{\theta(\lambda)}{2}d\lambda}\label{eq24}
	\end{equation}
	Differentiating both sides w.r.t $\lambda$ we get
	\begin{equation}
		\dfrac{d\sigma}{d\lambda}=\left(\frac{d\Sigma}{d\lambda}-\frac{\theta(\lambda)\Sigma(\lambda)}{2}\right)e^{-\int \frac{\theta(\lambda)}{2}d\lambda}\label{eq25}
	\end{equation}
	From equations (\ref{eq23}) and (\ref{eq24}) we have
	\begin{equation}
		\dfrac{d\sigma}{d\lambda}=-\theta(\lambda)\Sigma(\lambda)e^{-\int \frac{\theta(\lambda)}{2}d\lambda}+N\label{eq26}
	\end{equation}
	where $N$ represents the non-linear terms including the Weyl tensor that carries the degrees of freedom of the gravitational field e.g. GWs in weak-field limit (in the absence of $R_{ab}$). From (\ref{eq25}) and (\ref{eq26}) we have 
	\begin{eqnarray}
		\left(\dfrac{d\Sigma}{d\lambda}-\dfrac{\theta\Sigma}{2}\right)e^{-\int \frac{\theta(\lambda)}{2}d\lambda}=--\theta(\lambda)\Sigma(\lambda)e^{-\int \frac{\theta(\lambda)}{2}d\lambda}\nonumber\\+N\\
		\implies \dfrac{d\Sigma}{d\lambda}+\frac{\theta\Sigma}{2}=Ne^{\int \frac{\theta(\lambda)}{2}d\lambda}\label{eq28}
	\end{eqnarray}
	Differentiating both sides of equation (\ref{eq28}) we get,
	\begin{equation}
		\dfrac{d^{2}\Sigma}{d\lambda^{2}}+\dfrac{\theta}{2}\dfrac{d\Sigma}{d\lambda}+\dfrac{\Sigma}{2}\dfrac{d\theta}{d\lambda}=\left(\frac{dN}{d\lambda}+\frac{\theta(\lambda)}{2}\right)e^{\int \frac{\theta(\lambda)}{2}d\lambda}\label{eq29}
	\end{equation}
	The above equation resembles a damped harmonic oscillator equation in $\Sigma$ where $\theta$ )(expansion/contraction of the bundle) denotes the damping term, frequency is given by the rate of expansion $\frac{d\theta}{d\lambda}$ and the driving force comes from the Weyl contribution. Retaining $\theta$ in the shear evolution equation is therefore useful because it acts as a damping term when the equation is cast into the damped oscillator form. Thus the approximation $\theta \approx 0$ should be interpreted as $\theta^2 \ll \sigma^2$ rather than the strict condition $\theta = 0$. $\Sigma$ measures the shear amplitude which in the context of GWs is the wave amplitude.
	
	Shear encodes GW effects geometrically. Shear influences the distortion of a bundle of geodesics. These distortions are exactly done by a passing GW to a ring of test particles. Therefore, locally the effect of GWs is captured by shear via the shear equation. Moreover, in perturbation theory GW variables often suffer from gauge ambiguities  but shear $\sigma_{ab}$ being a covariant tensor (coordinate independent) make the evolution gauge-invariant. From the damped oscillator version of the shear equation it is clear that in Friedmann–Lemaître–Robertson–Walker (FLRW) universe (expanding universe) damping of GWs will be locally observed as the damping term is proportional to $\theta$ which is $3H$ (thrice the Hubble parameter in FLRW background). Thus, evolution of shear (one of the REs) describes GW amplitude evolution ($\Sigma$) in expanding or curved space-time. Most importantly, shear is driven by Weyl tensor which is precisely the carrier of GWs. This can be understood if we invoke the geodesic deviation equation in weak-field limit
	\begin{equation}
		\dfrac{d^{2}\xi^{a}}{d\lambda^{2}}=-R^{a}_{bcd}k^{b}k^{d}\xi^{c}\approx -C^{a}_{bcd}k^{b}k^{d}\xi^{c}\label{eq39}
	\end{equation}

		The vector $\xi^{a}$ appearing in equation (\ref{eq39}) is known as the Jacobi field and describes the infinitesimal separation between neighboring geodesics in a congruence. Physically, it represents the relative displacement between two nearby test particles moving along geodesics. In the weak-field vacuum limit the Riemann tensor is dominated by the Weyl tensor, so that the distortion of the geodesic congruence is primarily driven by the Weyl curvature associated with gravitational waves.

	The shear (via Weyl tensor) encodes the relative distortion in the null congruence. Therefore, by measuring the rate at which this distortion oscillates one effectively measures the local GW frequency. In other words, in the presence of GWs a ring of test particles get distorted and this is measured by the evolution of the Jacobi field. While, if we focus on weak lensing we can explain in a similar manner. Weak lensing phenomena also being affected by evolution of shear is associated with the evolution of Jacobi field i.e, the geodesic deviation equation. In the damping transformation (\ref{eq24}), $\Sigma(\lambda)$ can be a component of Jacobi field. Oscillations in the Jacobi field directly affects oscillations in the GW field. Jacobi field behaves like an oscillator driven by curvature via Weyl tensor. In LIGO \cite{LIGOScientific:2017vwq}, the arms of the interferometer act like a Jacobi field-small separations between nearby geodesics (mirrors). A passing GW changes the distance between mirrors via geodesic deviation equation. The GW signal is nothing but measurement of Jacobi-field oscillations induced by Weyl curvature tensor. GWs do not cause permanent expansion but induce oscillatory shear. Moreover, the shear evolution as damped oscillator is mathematically analogous to the quasinormal mode (QNM) equation, where GWs after a merger settle into a damped oscillatory phase known as the ringdown \cite{Konoplya:2024vuj}, \cite{Chakraborty:2025jto}, \cite{Rosato:2024arw}. QNMs are governed by perturbations in the curvature of spacetime, particularly the Weyl tensor in vacuum. Since the letter emphasizes how Weyl curvature drives shear hence the evolution of shear naturally encodes QNM signatures especially in the post-merger vacuum region. Further, the letter connects shear and Jacobi fields via geodesic deviation. This has a direct physical analog in LIGO-like detectors where QNMs distort the separation of test masses \cite{Huang:2025wom}. Thus, modeling the Jacobi field as a damped oscillator is conceptually the same as modeling the ringdown phase of a black hole.

		An interesting implication of the present formalism arises when the background spacetime is taken to be the Friedmann--Lema\^{i}tre--Robertson--Walker (FLRW) cosmological model. In such a spacetime the expansion scalar of the congruence is related to the Hubble parameter through
		\begin{equation}
			\theta = 3H .
		\end{equation}
		As a consequence, in the damped oscillator form of the shear evolution equation the damping term becomes directly proportional to the cosmic expansion rate. This implies that gravitational waves propagating through an expanding universe experience cosmological damping governed by the Hubble parameter. From the point of view of observations, this connection suggests that measurements of gravitational-wave amplitudes or weak-lensing shear patterns could in principle be interpreted through the evolution of shear in the Raychaudhuri framework. Although the present work does not attempt to extract specific cosmological parameters from observational data analysis, the formulation provides a geometric basis for connecting gravitational-wave propagation and weak-lensing observables with cosmological expansion.
	\section{Conclusion}
	The RE, mostly celebrated for its central role in the formulation of gravitational singularity theorems possesses far broader physical relevance, particularly in the analysis of geodesic congruences and their kinematic properties. This letter has explored the underrated yet profound utility of the RE especially its shear evolution component in studying two observationally significant phenomena: GWs and gravitational lensing, within the weak-field limit.
	
	In scenarios involving twist-free (vorticity-free) null geodesic congruences, which are relevant for both point-source light propagation and plane gravitational waves, the evolution of shear emerges as the dominant mechanism influencing the behavior of the geodesic bundle. For weak gravitational lensing, which is commonly observed in astrophysical surveys, the deflection and distortion of light are governed primarily by the shear term in the Raychaudhuri framework alongwith the expansion scalar playing a negligible role. Similarly, for GWs propagating through vacuum or nearly vacuum spacetimes, the Weyl tensor acting via shear encapsulates the tidal deformation of geodesic congruences thereby effectively capturing the gravitational wave amplitude and its local effects on matter.
	
	A key contribution of this letter is the formulation of the shear evolution equation into a form analogous to that of a damped harmonic oscillator, where the expansion scalar $\theta$ acts as a damping term, the rate of expansion ($\frac{d\theta}{d\lambda}$) determines the natural frequency, and the Weyl curvature functions as the driving force. This analogy not only enhances conceptual clarity but also opens the door to analytical methods from classical mechanics being applied to relativistic phenomena. It highlights how the oscillatory behavior of GWs and the subtle distortions in weak lensing can both be understood through the dynamics of shear. Doing so is comfortable in the sense that shear is such a  quantity that remains covariant and gauge-invariant making physically robust analysis.
	
	Moreover, the Newtonian analogue of the RE derived in the letter bridges the gap between classical gravitational theory and general relativity, offering intuitive insight into how shear and vorticity encode gravitational effects even in a non-relativistic setting. In particular, it shows that for irrotational flows, the gravitational potential can be fully described in terms of shear, emphasizing the geometric nature of gravity. This unified treatment of GWs and lensing via the Raychaudhuri formalism suggests deeper geometric and physical connections between phenomena traditionally studied separately. It reinforces the viewpoint that spacetime curvature especially its Weyl component not only governs the evolution of the universe on large scales but also leaves observable imprints through lensing patterns and GW signatures.
	
	In conclusion, the RE when interpreted beyond its conventional singularity-theorem context serves as a powerful geometric and physical tool for probing the dynamics of spacetime. This letter underscores the central role of shear evolution in this regard and encourages further exploration of Raychaudhuri-based methods in both theoretical investigations and observational interpretations of gravitational phenomena. Therefore, the Raychaudhuri formalism, when interpreted through the dynamics of shear, provides a unified geometric framework for understanding gravitational wave propagation and weak gravitational lensing within the standard theory of General Relativity. In the weak-field limit of General Relativity, where spacetime curvature is sufficiently small for linearized gravity to provide an accurate description, the curvature tensors are obtained in simplified forms much considerably and the Ricci tensor contribution vanishes in vacuum, allowing the Weyl tensor to emerge as the dominant curvature component governing the evolution of shear. On the contrary, extending the analysis to strong-field regimes such as the vicinity of black holes or neutron stars imposes considerable nonlinear couplings between expansion, shear and curvature terms in the Raychaudhuri equations. In such situations the resulting system of equations generally do not admit simple analytical solutions and often requires numerical integration of the geodesic congruence in highly curved space-times. This is why, the present letter restricts attention to the weak-field scenario relevant to weak gravitational lensing and gravitational waves observed far from their sources. A systematic investigation of the strong-field regime within the Raychaudhuri--shear framework would be an interesting direction for future work.
	
	Further, the analysis considers twist-free (vorticity-free) congruences. A natural extension would be to explore the impact of non-zero vorticity on the evolution of geodesics, particularly in rotating spacetimes (e.g., Kerr geometry). The formulation of the shear evolution equation as a damped oscillator invites direct comparison with the QNM description of perturbed black holes during the ringdown phase. Future work can focus on expressing QNM frequencies  in terms of shear dynamics and Weyl curvature, thereby offering a geometric and gauge-invariant perspective on ringdown signals observed in gravitational wave detectors.
	\section*{appendix}
		The connection between the Raychaudhuri equation and the classical Newtonian fluid equations can be made explicit in the weak-field and slow-motion limit. In this limit the four–velocity of the fluid may be written as
		\begin{equation}
			u^a \approx (1,\vec{v}),
		\end{equation}
		where $\vec{v}$ is the ordinary three–velocity. The expansion scalar then reduces to the divergence of the velocity field
		\begin{equation}
			\theta = \nabla_a u^a \approx \nabla \cdot \vec{v}.
		\end{equation}
		
		For a pressureless fluid with negligible vorticity, the time-like Raychaudhuri equation reduces to
		\begin{equation}
			\frac{d\theta}{dt} = -\frac{1}{3}\theta^2 - 2\sigma^2 - R_{ab}u^a u^b .
		\end{equation}
		
		In the Newtonian limit the curvature term can be expressed using the Poisson equation,
		\begin{equation}
			R_{ab}u^a u^b \approx \nabla^2 U = 4\pi G \rho .
		\end{equation}
		
		Substituting $\theta = \nabla \cdot \vec{v}$ we obtain
		\begin{equation}
			\frac{d}{dt}(\nabla \cdot \vec{v}) =
			-\frac{1}{3}(\nabla \cdot \vec{v})^2 - 2\sigma^2 - 4\pi G\rho .
		\end{equation}
		
		On the other hand, taking the divergence of the Newtonian Euler equation
		\begin{equation}
			\frac{d\vec{v}}{dt} = - \nabla U
		\end{equation}
		gives
		\begin{equation}
			\frac{d}{dt}(\nabla \cdot \vec{v}) = - \nabla^2 U .
		\end{equation}
		Using the Poisson equation $\nabla^2 U = 4\pi G\rho$, we recover the same gravitational source term appearing in the Raychaudhuri equation. Thus the Newtonian limit of the Raychaudhuri equation reproduces the evolution equation for the divergence of the velocity field obtained from the classical Euler–Poisson system, providing a geometric interpretation of Newtonian gravitational dynamics.
	\section*{acknowledgment}
	The authors thank the anonymous learned referees for their insightful comments and suggestions that improved the quality and visibility of the letter to a great extent. M.C thanks Inter University Center for Astronomy and Astrophysics (IUCAA), Pune, India for their warm hospitality and research facilities where a part of this work was carried out under the General Relativity On-Site Refresher Course 2025. S.C thanks IUCAA, Pune, India for their Associateship program. M.C thanks Techno India University, West Bengal and S.C thanks Brainware University, West Bengal for providing research facilities. 
	
\end{document}